\documentclass[11pt,a4paper]{article}

\usepackage{amsmath}
\usepackage{amssymb}

\usepackage{graphicx}
\usepackage{float}
\usepackage[topadjust=-1pt,captionskip=0pt]{subfig}
\usepackage[space]{cite}

\usepackage{color} 
\usepackage{eucal} 
\usepackage{tabularx}

\usepackage{enumerate}

\newcommand{\me}[1]{eq.~{\eqref{eq:#1}}}
\newcommand{\mes}[1]{eqs.~{\eqref{eq:#1}}}
\renewcommand{\v}[1]{\boldsymbol{#1}}

\newcommand{\by}{\v y}
\newcommand{\bY}{\v Y}
\newcommand{\bZ}{\v Z}

\newcommand{\bW}{\v W}
\newcommand{\bWo}{\v W^{\text{old}}}
\newcommand{\bz}{\v z}
\newcommand{\R}{\mathbb{R}}

\newcommand{\MD}{\mathcal{M}}
\renewcommand{\P}{\mathcal{P}}

\usepackage{setspace} 
\title{A probabilistic model for learning in cortical microcircuit motifs with data-based divisive inhibition}
\author
{Robert Legenstein$^*$, Zeno Jonke$^*$, Stefan Habenschuss, Wolfgang Maass\\
\\
\normalsize{Institute for Theoretical Computer Science}\\
\normalsize{Graz University of Technology}\\
\normalsize{\today}\\
}

\date{}

\begin{document}

\maketitle
$^*$ These authors contributed equally to the work.

\vspace{5cm}

\begin{abstract}
Previous theoretical studies on the interaction of excitatory and inhibitory neurons proposed to model this cortical microcircuit motif as a so-called Winner-Take-All (WTA) circuit. A recent modeling study however found that the WTA model is not adequate for data-based softer forms of divisive inhibition as found in a microcircuit motif in cortical layer 2/3. 
We investigate here through theoretical analysis the role of such softer divisive inhibition for the emergence of computational operations and neural codes under spike-timing dependent plasticity (STDP).
We show that in contrast to WTA models --- where the network activity has been interpreted as probabilistic inference in a generative mixture distribution --- this network dynamics approximates inference in a noisy-OR-like generative model that explains the network input based on multiple hidden causes. 
Furthermore, we show that STDP optimizes the parameters of this model by approximating online the
expectation maximization (EM) algorithm. 
This theoretical analysis corroborates a preceding modelling study which suggested 
 that the learning dynamics of this layer 2/3 microcircuit motif extracts a specific modular representation of the input and thus performs blind source separation on the input statistics.
\end{abstract}

\newpage

\section{Introduction}

Winner-take-all-like (WTA-like) circuits constitute a ubiquitous motif of cortical microcircuits \cite{DouglasMartin:04}. Previous models and theories for competitive Hebbian learning in WTA-like circuit from \cite{RumelhartZisper:85} to \cite{nessler2013bayesian} were based on the assumption of strong WTA-like lateral inhibition. 
Several theoretical studies showed that spike-timing dependent plasticity (STDP) supports the emergence of Bayesian computation in such winner-take-all (WTA) circuits \cite{nessler2013bayesian, habenschuss2013emergence, KlampflETAL:13}. These analyses were based on a probabilistic generative model approach. In particular, it was shown that the network implicitly represents the distribution of input patterns through a generative mixture distribution and that STDP optimizes the parameters of this mixture distribution.  
But this analysis assumed that the input to a WTA is explained at any point in time by a single neuron, and that strong lateral inhibition among pyramidal cells ensures a basically fixed total output rate of the WTA. These assumptions, however, may not be suitable in the context of more realistic activity dynamics in cortical networks. 

In fact, recent modeling results \cite{avermann2012microcircuits, JonkeETAL:17a} show that the WTA model is not adequate for a softer form of inhibition that has been reported for cortical layer 2/3. This softer form of inhibition is often referred to as feedback inhibition, or lateral inhibition, and has been termed more abstractly based on its influence on pyramidal cells as divisive inhibition \cite{WilsonETAL:12,CarandiniHeeger:12}. It stems from dense bidirectional interconnections between layer 2/3 pyramidal cells and nearby Parvalbumin-positive ($\text{PV}^+$) interneurons (often characterized as fast-spiking interneurons, in particular basket cells), see e.g. \cite{PackerYuste:11, FinoETAL:12,avermann2012microcircuits}. 
The simulations results in \cite{JonkeETAL:17a} also indicate that blind source separation emerges as the computational function of this microcircuit motif when STDP is applied to the input synapses of the circuit.

The results of \cite{JonkeETAL:17a} raise the question whether they can be understood from the perspective of a corresponding probabilistic generative model, that could replace the mixture model that underlies the analysis of emergent computational properties of microcircuit motivs with hard WTA-like inhibition. We propose here such a model that is based on a Gaussian prior over the number of active excitatory neurons in the network and a noisy-OR-like likelihood term. We develop a novel analysis technique based on the neural sampling theory \cite{BuesingETAL:11} to show that the microcircuit motif model approximates probabilistic inference in this probabilistic generative model. Further, we derive a plasticity rule that optimizes the parameters of this generative model through online expectation maximization (EM), the arguably most powerful tool from statistical learning theory for the optimization of generative models. We show that this plasticity rule can be approximated by an STDP-like learning rule.

This theoretical analysis strengthens the claim that blind source separation \cite{foldiak1990forming} --- also referred to as independent component analysis \cite{HyvaerinenETAL:04} --- emerges as 
a fundamental computation on assembly codes through STDP in this microcircuit motif. This computational operation enables a network to disentangle and separately represent superimposed inputs that result from independent assembly activations in different upstream networks. Furthermore, our theoretical analysis reveals that the ability of this cortical microcircuit motif to perform blind source separation is facilitated either by the normalization of activity patterns in input populations, or by homeostatic mechanisms that normalize excitatory synaptic efficacies within each neuron.

\begin{figure}[t]
\begin{center}
\includegraphics[width=0.8\textwidth]{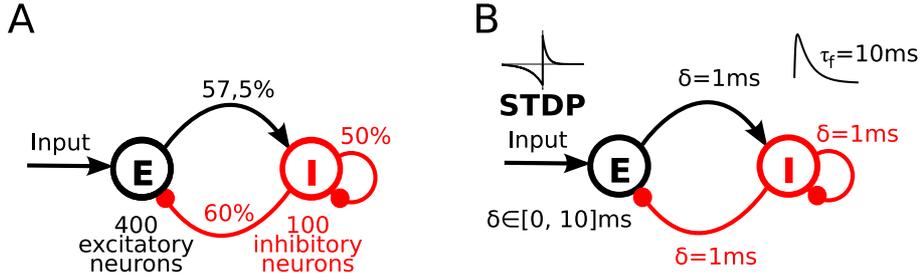}
\end{center}
\caption{{\bf Data-based network model $\MD$ for a microcircuit motif.}
\textbf{A}) Network anatomy. Circles denote excitatory (black) and inhibitory (red) pools of neurons. Black arrows indicate excitatory connections. Red lines with dots indicate inhibitory connections. Numbers above connections denote corresponding connection probabilities. \textbf{B}) Network physiology. Same as in (A), but connection delays $\delta$ are indicated. All synapses are modeled with the same PSP shape using a decay time constant of $\tau_\text{f}=10$ ms as indicated on top right. Input synapses are subject to STDP. }
\label{fig:fig_model}
\end{figure}

\section{Results}

A data-based microcircuit motif model for the interaction of pyramidal cells with $\text{PV}^+$ inhibitory neurons in layer 2/3 has been introduced in \cite{avermann2012microcircuits}. Based on this study, \cite{JonkeETAL:17a} analyzed the computational properties that emerge in this microcircuit motif from synaptic plasticity.  We first briefly introduce the microcircuit motif model analyzed in \cite{JonkeETAL:17a} and discuss its properties. Subsequently, we present a theoretical analysis of this network motif based on a probabilistic generative model $\mathcal{P}$.

\subsection{A data-based model for a network motif consisting of excitatory and inhibitory neurons}\label{sec:ei-model}
\cite{JonkeETAL:17a} proposed a specific model for interacting populations of pyramidal cells with $\text{PV}^+$ inhibitory neurons in cortical layer 2/3  based on data from the Petersen Lab \cite{avermann2012microcircuits}, see Fig.~\ref{fig:fig_model}A, B. We refer to this specific model as the microcircuit motif model $\MD$.

The model $\MD$ consists of
two reciprocally connected pools of neurons, an excitatory pool and an inhibitory pool.  
$M$ stochastic spiking neurons constitute the excitatory pool. Their dynamics is given by a stochastic version of the spike response model that has been fitted to experimental data in \cite{JolivetETAL:06}.
The instantaneous firing rate $\rho_m$ of a neuron $m$ depends exponentially on its current membrane potential $u_m$,
\begin{align}\label{eq:neuron_firing_prob}
  \rho_m(t)\ &= \frac{1}{\tau} \exp(\gamma \cdot u_m(t))\;,
\end{align}
where $\tau=10$ ms and $\gamma=2$ are scaling parameters that control the shape of the response function.
After emitting a spike, the neuron enters a refractory period.

The excitatory neurons are reciprocally connected to a pool of recurrently connected inhibitory neurons. All connection probabilities in the model were taken from \cite{avermann2012microcircuits}.
Excitatory neurons receive excitatory synaptic inputs $\tilde y_1(t),..,\tilde y_N(t)$ with corresponding synaptic efficiencies $w_{im}$ between the input neuron $i$ and neuron $m$. These afferent connections are subject to a standard form of STDP.
Thus, the membrane potential of excitatory neuron $m$ is given by the sum of external inputs, inhibition from inhibitory neurons, and its excitability $\alpha$
\begin{align}
u_m (t) &= \sum_i w_{im} \tilde y_i(t) -\sum_{j \in \mathcal{I}_m} w^{\text{IE}} I_j(t) + \alpha,\label{eq:membrane}
\end{align}
where $\mathcal{I}_m$ denotes the set of indices of inhibitory neurons that project to neuron $m$, and $w^{\text{IE}}$ denotes the weight of these inhibitory synapses. $I_j(t)$ and $\tilde y_i(t)$ denote synaptic input from inhibitory neurons and input neurons respectively, see above.

Inhibitory contributions to the membrane potential of pyramidal cells have in this neuron model a divisive effect on the firing rate. This can be seen by by substituting \me{membrane} in \me{neuron_firing_prob}, see also eq.~\eqref{eq:neuron_firing_prob_divisive} in {\em Methods}, thus implementing divisive inhibition (see \cite{CarandiniHeeger:12} for a recent review).
Divisive inhibition has been shown to be 
a ubiquitous computational primitive in many brain circuits (see \cite{CarandiniHeeger:12} for a recent review). In mouse visual cortex, divisive inhibition is implemented through $\text{PV}^+$ inhibitory neurons \cite{WilsonETAL:12}. 
Although the inhibitory signal is common to all neurons in the pool of excitatory neurons, contrary to the inhibition modeled in \cite{nessler2013bayesian} it does not normalize the firing rates of neurons exactly and therefore the total firing rate in the excitatory pool is variable and depends on the input strength. Importantly, in contrast to \cite{nessler2013bayesian} where inhibition strictly enforced that only a single neuron in the excitatory pool is active at any given time, the data-based model $\MD$ allows several neurons to be active concurrently. 

\begin{figure}
\begin{center}
\includegraphics[width=0.85\textwidth]{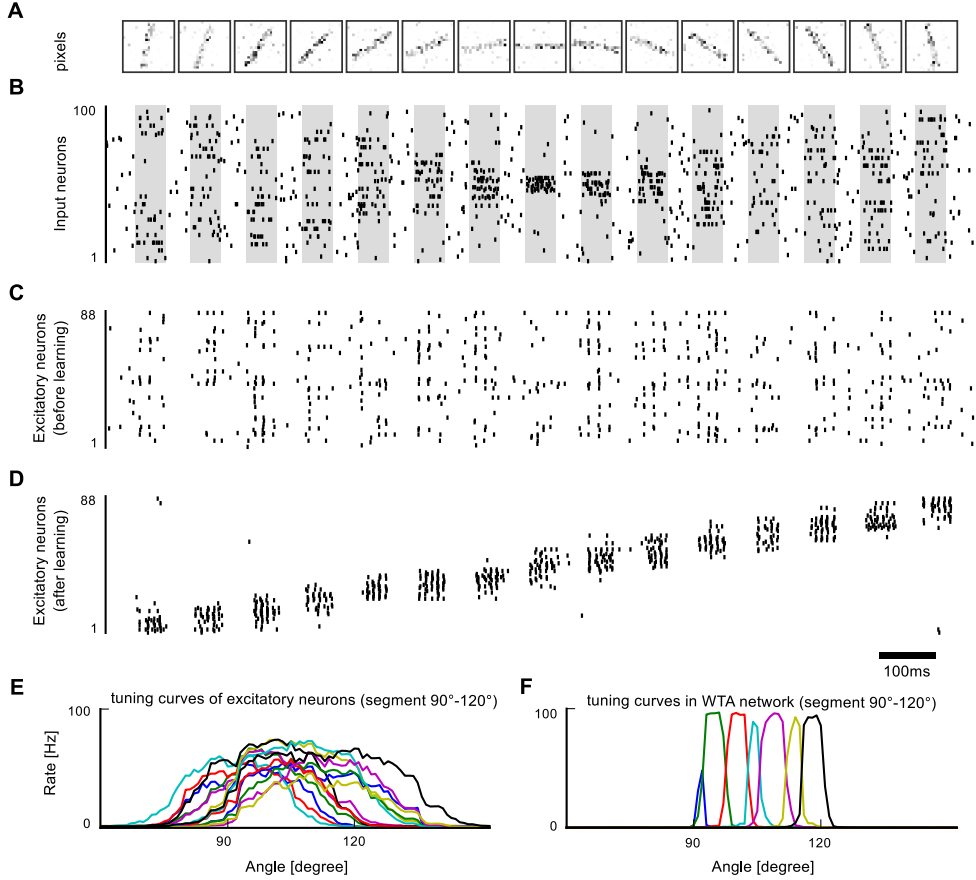}
\end{center}
\caption[]{{\bf Emergent computational properties of the data-based network model $\MD$.} {\bf A}) Network inputs are given by images of randomly oriented bars (inputs arranged in 2D for clarity; pixel gray-level indicates effective network input $\tilde y_i(t)$, see eq.~(\ref{eq:output_trace})). {\bf B}) Input neuron spike patterns (every $4^{th}$ neuron shown). Presence of a bar in the input with orientation indicated in panel A indicated by gray shading. {\bf C, D}) Spike responses of a subset of excitatory neurons in $\MD$ to the input in (B) before (C) and after (D) learning (neurons sorted by preferred orientation). {\bf E}) Tuning curves of excitatory neurons in with preferred orientations between $90$ and $120$ degrees. {\bf F}) Orientation tuning curves in a WTA model \cite{nessler2013bayesian,habenschuss2013emergence}. Figure modified from Fig.~2 in \cite{JonkeETAL:17a}.
}
\label{fig:fig_tilted}
\end{figure}

\subsection{Emergent properties of the data-based network model: From WTA to k-WTA}\label{sec:emergent-coding-properties}

The computational properties of this data-based network model $\MD$  were extensively studied through simulations in \cite{JonkeETAL:17a}.
In order to compare the properties of this data-based network model $\MD$ to previously considered WTA models, they examined the emergence of orientation selectivity, which we briefly discuss here. For details, please see \cite{JonkeETAL:17a}.  Pixel-representations of noisy bars in random orientations were provided as external spike inputs
 (Fig.~\ref{fig:fig_tilted}A). Input spike trains were generated from these pixel arrays by converting pixel values to Poisson firing rates of input neurons (black pixel: $75$ Hz; white pixel: $1$ Hz). Randomly oriented bars were presented to the network for $400$ s where each bar was presented for $50$ ms, see Fig.~\ref{fig:fig_tilted}B (STDP was applied to synapses from input neurons to excitatory neurons). The resulting network response (Fig.~\ref{fig:fig_tilted}D) 
shows the emergence of assembly codes for oriented bars. The resulting Gaussian-like tuning curves of excitatory neurons (Fig.~\ref{fig:fig_tilted}E) densely cover all orientations, resembling experimental data from orientation pinwheels (see Fig. 2 d,e in \cite{OhkiETAL:06}). Also consistent with experimental data \cite{KerlinETAL:10,IsaacsonScanziani:11}, inhibitory neurons did not exhibit orientation selectivity (not shown).

In contrast, previously considered models with idealized strong inhibition in WTA-circuits \cite{nessler2013bayesian} show a clearly distinct behavior, see Fig.~\ref{fig:fig_tilted}F. For this model, at most a single neuron could fire at any moment of time, and as a result at most two neurons responded after a corresponding learning protocol with an increased firing rate to a given orientation (see Fig.~\ref{fig:fig_tilted}F and Fig.~5 in \cite{nessler2013bayesian}).
In the simulations of the data-based model $\MD$, on average $k=17$ neurons responded to each orientation with an increased firing rate. This suggests that the emergent computational operation of the layer 2/3 microcircuit motif with divisive inhibition is better described as k-WTA computation, where $k$ winners may emerge simultaneously from the competition. This number $k$ is however not a strict constraint in the data-based model $\MD$. The actual number of winners depends on synaptic weights and the external input. Its computation is thus better describe as an adaptive k-WTA operation.  
The k-WTA characterisitcs of the layer 2/3 microcircuit motif is quite attractive, since it is known from computational complexity theory that the k-WTA computation is more powerful than the simple WTA computation (for $k > 1$) \cite{Maass:00}.

\subsection{Theoretical framework for understanding emergent computational properties the layer 2/3 microcircuit motif}\label{sec:theory}

Fig.~\ref{fig:fig_tilted} demonstrates that significantly different computational properties emerge in the data-based model $\MD$ through STDP as compared to previously considered WTA models \cite{nessler2013bayesian,habenschuss2013emergence}. 
The main aim of this article is to understand this different emergent computational capability theoretically, in particular since the analysis from \cite{nessler2013bayesian} and \cite{habenschuss2013emergence} in terms of mixture distributions is only applicable to WTA circuits. 
The novel analysis technique that we will use is summarized as follows. First, using some simplifications on the network dynamics, we formulate the network dynamics in the neural sampling framework  \cite{BuesingETAL:11}. This allows us to deduce the distribution $p(\v z | \v y, \v W)$ of activities $\v z$ of excitatory neurons in the network for a given input $\v y$ and for the given network weights $\v W$. We then show that this distribution approximates the posterior distribution of a generative probabilistic model $\P$. 
This generative model is not a mixture distribution as in the WTA case  \cite{nessler2013bayesian}, but a more complex distribution that is based on a noisy-OR-like likelihood.
We make the nature of the approximation explicit and evaluate its severity through simulations. Finally, we derive a plasticity rule that implement online EM in this generative model, thus implementing blind source separation. We find that this plasticity rule can be approximated by an STDP-like learning rule.

\subsubsection{Formulation of the network dynamics of $\MD$ in the neural sampling framework}
The neural sampling framework \cite{BuesingETAL:11} provides us with the ability to determine the stationary distribution (defined in the following) of network states for the given network parameters and a given network input. 
In order to be able to describe the probabilistic relationships between input and network activity, we 
describe network inputs by binary vectors $\by(t)$ and responses of excitatory neurons in $\MD$ by binary vectors $\bz(t)$.
The vectors $\by(t)$ and $\bz(t)$ capture the spiking activity of ensembles of spiking neurons in continuous time according to the common convention introduced in \cite{BerkesETAL:11} and \cite{BuesingETAL:11}: A spike of the $i^{th}$  neuron in the ensemble at time $t$ sets the corresponding component $y_i(t)$ ($z_i(t)$) of the bit vector $\by(t)$ ($\bz(t)$) from its default value $0$ to $1$ for some duration $\tau$ (that can be chosen for example to reflect the typical time constant of an EPSP), see {\em Methods}. Note the difference between the vectors $\by(t)$, $\bz(t)$ and the output traces $\tilde \by(t)$, $\tilde \bz(t)$ used in eq.~\eqref{eq:membrane} (and defined in eq.~\eqref{eq:output_trace} in {\em Methods}). The former constitute an abstract convention to describe the momentary state of the network based on its current firing activity, while the latter describe the impact that the neurons have on their postsynaptic targets in terms of real-valued double-exponential EPSPs.

We want to describe the distribution of network states $\bz(t)$ for given inputs $\by(t)$ and network parameters $\bW$ in terms of a probability distribution $p(\bz | \by, \bW)$.  
In this distribution, the activities of network inputs and excitatory neurons in $\MD$ are represented by two vectors of binary random variables: $\v y = (y_1, \dots, y_N)$ (termed input variables in the following) and $\v z = (z_1, \dots, z_M)$ (termed hidden variables or hidden causes).   
The network state $\by(t), \bz(t)$ at time $t$ is interpreted as one specific realization of these random variables. 

In order to make this mapping between network activity in $\MD$ and the distribution of network states feasible, one has to make three simplifying assumptions about the dynamics of the neural network models similar as in \cite{BuesingETAL:11}. 
First, PSPs of inputs and network neurons are rectangular with length $\tau$ (chosen here to be $\tau=10$ ms) and network neurons are refractory for the same time span $\tau$ after each spike. Second, synaptic connections are idealized in the sense that the synaptic delay is $0$ (i.e., a presynaptic spike leads instantaneously to a PSP in the postsynaptic neuron). And finally, the weights of recurrent synaptic connections are symmetric (i.e., the weight from neuron $i$ to neuron $j$ is identical to the weight from neuron $j$ to neuron $i$). This necessitates that lateral inhibition is not implemented through a pool of inhibitory neurons. 
Instead, the network dynamics is defined by only one pool of $M$ network neurons (the same number as the number of excitatory network neurons in $\MD$). Since the inhibitory neurons in $\MD$ show linear response properties, the inhibition in the network depends linearly on the activity of excitatory neurons in the network. 
One can therefore model the inhibition in the network by direct inhibitory connections between excitatory neurons (where synaptic delays are neglected) with weight $\beta$. 
For clarity, we provide the full description of the approximate dynamics in the following. 

The approximate dynamics is described by $M$ network neurons. 
Network neurons have instantaneous firing rates that depend exponentially on their membrane potential, as given in \me{neuron_firing_prob}. Whenever neuron $m$ spikes, the output trace $\tilde z_m(t)$ of neuron $m$ is set to $1$ for a period of duration $\tau$ (this corresponds to a rectangular PSP; the same definition applies to output traces $\tilde y_m(t)$ of input neurons). 
After emitting a spike, the neuron enters a refractory period of duration $\tau=10$ ms, during which its instantaneous spiking probability is zero.
Note that for this definition of the output trace, the state vector $\bz(t)$ is identical to the vector of output traces $\tilde \bz(t)$.
Lateral inhibition in the network is established by direct inhibitory connections between excitatory neurons, leading to membrane potentials
\begin{equation}\label{eq:membrane_NS}
u_m(t) = \sum_{i}^N w_{im} \tilde y_i(t) - \sum_{j\neq m} \beta \tilde z_j(t) +  c_m ,
\end{equation}
where $c_m$ denotes some neuron-specific excitability of the neuron that is independent of the input and network activity. 
Each network neuron receives feedforward synaptic inputs $\tilde y_1(t),\dots ,\tilde y_N(t)$ whose contribution to the membrane potential of a neuron $m$ at time $t$ depends on the synaptic efficiency $w_{im}$ between the input neuron $i$ and the network neuron $m$. Network neurons are all-to-all recurrently connected. 
The second term in \eqref{eq:membrane_NS} specifies this recurrent input where $\beta$ is the inhibitory recurrent weight of the connection between network neuron $j$ and network neuron $m$. 
It has been shown in \cite{BuesingETAL:11} that for such membrane potentials, the distribution of network states is given by the Boltzmann distribution:
\begin{align}\label{eq:post_NS}
p_\text{Network}(\v z| \v y, \v W) = \frac{1}{Z}\exp\left\{ \gamma \cdot \left( \sum_{i,m} w_{im} y_i z_m + \frac{1}{2}\sum_{m\neq l} \beta z_m z_l + \sum_m c_m z_m \right) \right\} ,
\end{align}
where $Z$ is a normalizing constant.

\begin{figure}
\begin{center}
\includegraphics[width=0.8\textwidth]{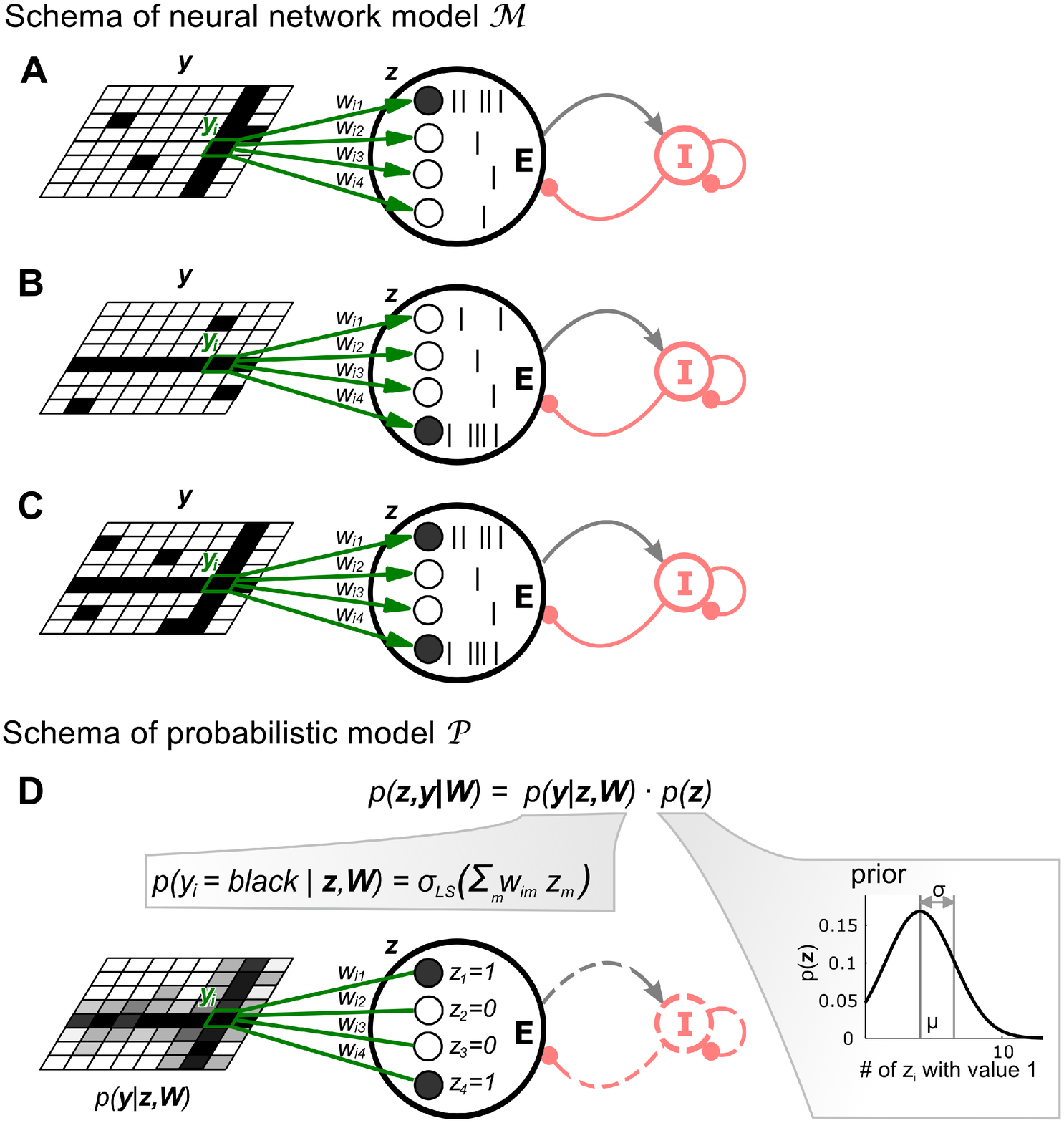}
\end{center}
\caption{{\bf Relationship between the data-based model $\boldsymbol{\MD}$ and the generative probabilistic model $\boldsymbol{\P}$.} 
{\bf A-C}) Schema of the response of model $\MD$ to superimposed bars. Network inputs (left) are schematically arranged in a 2D array for clarity of the argument. Black indicates highly active inputs neurons. {\bf A}) A vertical bar with added noise is presented to $\MD$. This input activates an excitatory neuron (filled circle, spiking activity is indicated), similar as in hard WTA models. {\bf B}) Another neuron is activated by a horizontal bar, also similar as in hard WTA models. {\bf C}) A combination of these two basic input patterns activates both neurons in $\MD$. This response is inconsistent with a WTA model, and with any generative model based on mixture distributions. But it can still be viewed as approximate inference of the posterior distribution $p(\bz|\by,\bW)$ over hidden causes $\bz$ for the given inputs $\by$ in the probabilistic model $\P$ shown in D. {\bf D}) Schema of probabilistic model $\P$. The joint $p(\bz, \by | \bW)$ is defined by the prior $p(\bz)$ and the likelihood $p(\by|\bz,\bW)$. Synaptic efficacies $\bW$ implicitly define the likelihood over inputs $\by$ for given hidden causes $\bz$ (probability values for inputs $y_i$ indicated by shading of squares; $\sigma_\text{LS}$ denotes the logistic sigmoid function). In the likelihood model, 
a given input $y_i$ is $1$ (corresponding to a black pixel in this example) with high probability if it has a large $w_{im}$ to at least one active hidden cause $z_m$. In the depicted example, $y_i$ belongs to two bars (see A, B) with corresponding active hidden causes. Due to the nonlinear behavior of the likelihood, its probability is comparable to one where only one of the hidden causes $z_m$ is active. The inset on the right depicts the Gaussian prior $p(\bz)$ on hidden causes $\bz$ with $\mu=4$ and $\sigma=2.5$. The prior implicitly incorporates in $\P$ the impact of the inhibitory feedback in the data-based model $\MD$ (therefore indicated with dashed lines).} 
\label{fig:fig_generative_model}
\end{figure}

\subsubsection{A probabilistic model $\boldsymbol{\P}$ for the layer 2/3 microcircuit motif:} 

Fig.~\ref{fig:fig_generative_model}A-C illustrates the putative stochastic computation performed by the model $\MD$, i.e., how network input leads to network activity in the model. 
Assume that we have a probabilistic model $\P$ for the inputs $\by$ defined by a prior $p(\bz)$ and a likelihood $p(\by|\bz, \bW)$. These distributions describe how one can generate input samples $\by$ by first drawing a {\em hidden state} vector $\bz$ from $p(\bz)$
and then drawing an input vector $\by$ from $p(\by|\bz, \bW)$. Therefore, such a probabilistic model $\P$ is also called a generative model (for the inputs).

If the distribution of network states \eqref{eq:post_NS} is the posterior distribution given by 
\begin{align}\label{eq:posterior-spelled}
p(\bz| \by, \bW) = \frac{p(\v z) p(\v y| \v z, \v W)}{\sum_{\bz'} p(\v y, \v z'|\v W)},
\end{align}
then the network performs probabilistic inference in this probabilistic model $\P$.
The inference task described by eq.~\eqref{eq:posterior-spelled} 
 assumes that $\by$ is given and the hidden causes $\bz$ (such as the basic components of a visual scene) have to be inferred. This inference can intuitively also be described as providing an "explanation" $\bz$ for the current observation $\by$ according to the generative model $\P$.
In the following, we describe a probabilistic model $\P$ and show that eq.~\eqref{eq:post_NS} approximates the posterior distribution of this model. This implies that the simplified dynamics of the data-based model $\MD$  approximate  probabilistic inference in the probabilistic model $\P$.

The probabilistic model $\P$ is defined by two distributions, the prior over hidden variables $p(\bz)$ (that captures constraints on network activity imposed for example by lateral inhibition) and the conditional likelihood distribution over input variables $p(\by|\bz, \bW)$ that describes the probability of input $\by$ for a given network state $\bz$ in a network with parameters $\bW$. These two distributions define the joint distribution over hidden and visible variables since $p(\bz, \by | \bW) =  p(\by|\bz, \bW) p(\bz)$, see Fig.~\ref{fig:fig_generative_model}D.   
The specific forms of these two distributions in the probabilistic model $\P$ considered for $\MD$ are discussed in the following and defined by \mes{genmod_prior}-\eqref{eq:genmod_likefact} below.

In previously considered hard WTA models \cite{nessler2013bayesian,habenschuss2013emergence}, strong lateral inhibition was assumed. This corresponded to a prior where only a single component $z_j$ of the hidden vector $\bz$ can be active at any time. The biologically more realistic divisive inhibition in $\MD$ allows several of them to fire simultaneously. 
This corresponds to a prior that induces sparse activity
in a soft manner (``adaptive'' k-WTA): It does not enforce a strict ceiling $k$ on the number of $z$-neurons that can fire within a time interval of length $\tau$, but only tries to keep this number within a desired range. 
Hence we use as prior in $\P$ a Gaussian distribution
\begin{align}\label{eq:genmod_prior}
p(\v z) &= \frac{1}{Z_{\text{prior}}} \exp\left(- \frac{1}{2\sigma^2}\left(\sum_{m=1}^M z_m - \mu\right)^2\, \right)\quad ,
\end{align}
where $Z_{\text{prior}}$ is a normalizing constant, $\mu \in \R$ is a parameter that shifts the mean of the distribution, and $\sigma^2>0$ defines the variance (see Fig.~\ref{fig:fig_generative_model}D). Note that the Gaussian is 
restricted to integers as the sum runs over binary random variables $z_1, \ldots, z_M$.

\begin{figure}[t]
\begin{center}
\includegraphics[width=0.9\textwidth]{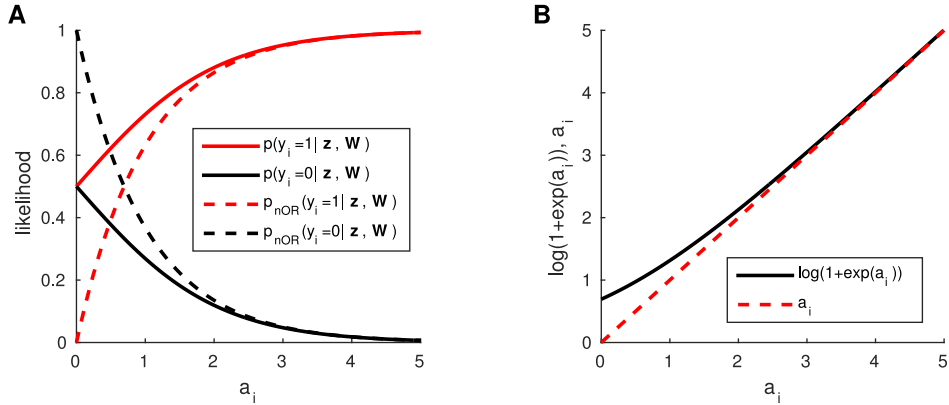}
\end{center}
\caption{{\bf Likelihood model of $\P$. (A)}
Likelihood that an input $y_i$ is $1$ (red) or $0$ (black) for the proposed likelihood model \eqref{eq:genmod_likefact} (full lines) and for the noisy-OR likelihood \eqref{eq:nOR} (broken lines). The likelihood of an input $y_i$ depends on the hidden causes $\v z$ through the weighted contribution $a_i =  \gamma \bar{\v w}_i^T \v z$ of the hidden causes to this input. For large $a_i$, the likelihood of $y_i=1$ approaches $1$. While for $a_i=0$, the likelihood of $y_i=1$ is $0.5$ in the proposed likelihood model, whereas it is $0$ in the noisy-OR model. {\bf B} Approximation of $\log (1+\exp(a_i))$ (black full line) by $a_i$ (red broken line) as used in Eq.~\eqref{eq:appr_A}.
}
\label{fig:likelihood}
\end{figure}

As in 
other generative models
we assume for the sake of theoretical tractability that the conditional likelihood $p(\v y|\v z, \v W)$ factorizes, so that each input $y_i$ is independently explained by the current network state $\bz$:
\begin{align}\label{eq:genmod_condlike}
p(\v y|\v z, \v W) &= \prod_{i=1}^N p(y_i|\v z, \v W).
\end{align}
A probabilistic model for hard WTA circuits can only explain each input variable $y_i$ by a single hidden cause $z_m$. 
In contrast, in probabilistic models with soft inhibition and the prior \eqref{eq:genmod_prior}, several hidden causes can be active simultaneously and explain together an input variable $y_i$. We define $\bar{\v w}_i$ as the vector of weights $\bar{\v w}_i = (w_{i1}, \dots, w_{iM})^T$ that define the likelihood for variable $y_i$. We consider the following likelihood model
\begin{align}
p(y_i|\v z, \v W) &= \frac{\exp({\gamma \bar{\v w}_i^T} \v z)^{y_i}}{1+\exp({\gamma \bar{\v w}_i^T} \v z)} = \frac{\exp(a_i)^{y_i}}{1+\exp(a_i)} = \sigma_\text{LS}\left((-1)^{y_i} a_i\right), \label{eq:genmod_likefact}
\end{align}
where we have defined $a_i =  \gamma \bar{\v w}_i^T \v z$, and $\sigma_\text{LS}$ is the logistic sigmoid $\sigma_\text{LS}(u) = \frac{1}{1+\exp(-u)}$. This likelihood function is shown in Figure \ref{fig:likelihood}A together with the often used noisy-OR likelihood. Note that if none of the hidden causes $z_m$ is active, i.e., $z_m=0$ for all $m$, then $y_i=1$ with probability $0.5$. Each active hidden cause $z_m=1$  with $w_{im}>0$ increases the probability that input variable $y_i$ assumes the value $1$, see also Fig.~\ref{fig:fig_generative_model}D). This likelihood, allows the generative model to deal with situations where an input neuron can fire in the context of different hidden causes, for example with pixels in the network inputs that lie in the intersection of different patterns, see Fig.~\ref{fig:fig_generative_model}). The soft Gaussian prior \eqref{eq:genmod_prior} allows the internal model to develop modular representations for different components of complex input patterns.
This likelihood is quite similar to the frequently used noisy-OR model (see e.g. \cite{neal1992connectionist,saund1995multiple}): 
\begin{align}
p_\text{nOR}(y_i=0|\v z, \v W) &= \exp(-a_i)^{1-y_i}(1-\exp(-a_i))^{y_i} \label{eq:nOR}
\end{align}
One difference is that (for purely excitatory weights), the probability of an input $y_i$ being zero is at most $0.5$ in the proposed likelihood, while it can become $0$ in the noisy-OR model. Such a model may reflect the situation that network inputs are noisy, so their firing rates are never zero.

We now analyze the relationship between the probabilistic model $\P$ and the description of the data-based model $\MD$ in the neural sampling framework. We will see that $\MD$ approximates probabilistic inference in $\P$. Finally, we show that adaptation of network parameters in $\MD$ through STDP can be understood as an approximate stochastic expectation maximization (EM) process in the corresponding probabilistic model $\P$. 

\subsubsection{Interpretation of the dynamics of $\boldsymbol{\MD}$ in the light of $\boldsymbol{\P}$: }


Using the likelihood and prior of $\P$, the posterior of hidden states $\v z$ for given inputs $\v y$ and given parameters $\v W$ is 
\begin{align}\label{eq:post}
p(\v z| \v y, \v W) = \frac{1}{Z}\exp\left\{ \gamma \cdot \left(\sum_{i,m} w_{im} y_i z_m - \frac{1}{2} \sum_{m\neq l} \beta z_m z_l + \sum_m \alpha z_m - \frac{1}{\gamma} \sum_i \log (1+\exp(a_i))\right) \right\},
\end{align}
where $Z$ is a normalizing constant, $\beta=\frac{1}{\gamma \sigma^2}$, and $\alpha=\frac{2\mu-1}{2 \gamma \sigma^2}$. The terms including $\beta$ and $\alpha$ stem from the prior, while the last term stems from the normalization of the likelihood. When we compare this posterior to the posterior of the network model eq.~\eqref{eq:post_NS}, we see that they are quite similar with $\beta$ denoting the strength of inhibitory connections and $\alpha$ being the neural excitabilities.

The last term in eq.~\eqref{eq:post} is problematic since the $a_i$'s depend on $\v z$ and thus the whole posterior is not a Boltzmann distribution and can therefore not be computed by the model $\MD$. It turns out however that this last term can be approximated quite well by a term that is linear in $\v z$. 
Note that for zero $a_i$ (i.e., for zero weights or for the zero-$\v z$-vector), this last term evaluates to $\log(2)$. But as $a_i$ increases, one can neglect the $1$ in the logarithm and the expression quickly approaches $a_i$. We can thus write
\begin{align}\label{eq:appr_A}
\sum_i \log (1+\exp(a_i)) \approx \sum_i a_i = \gamma \sum_i \sum_m w_{im} z_m = \gamma \sum_m z_m \left(\sum_i w_{im}\right),
\end{align}
where the term in the brackets on the right is just the L1-norm of the weight vector of neuron $j$. Note that for a given weight matrix, an increased $\gamma$ leads to a better approximation. The approximation of $\log (1+\exp(a_i))$ by $a_i$ is illustrated in Fig.~\ref{fig:likelihood}B.
Hence, the first approximate posterior we consider is given by 
\begin{align}\label{eq:post_A1}
p_\text{A1}(\v z| \v y, \v W) = \frac{1}{Z}\exp\left\{ \gamma \cdot \left( \sum_{i,m} w_{im} y_i z_m - \frac{1}{2} \sum_{m\neq l} \beta z_m z_l + \sum_m \alpha z_m - \sum_m z_m \sum_i w_{im}\right)    \right\}.
\end{align}
This is a Boltzmann distribution of the form \eqref{eq:post_NS} and the last term accounts to a neuron-specific homeostatic bias that depends on the sum of incoming excitatory weights. 
Matching the terms of this equation to the terms in eq.~\eqref{eq:post_NS} and performing the same match in the membrane potential \eqref{eq:membrane_NS}, we see that the membrane potential of neurons in this approximation is given by
\begin{equation}\label{eq:membrane_A1}
u_m(t) = \sum_{i}^N w_{im} \tilde y_i(t) - \sum_{j\neq m} \beta \tilde z_j(t) +  \alpha - \sum_i w_{im}.
\end{equation}
If excitatory weight vectors are normalized to an L1-norm of $w_\text{norm}=\sum_i w_{im}$ for all $m$, this simplifies to 
\begin{equation}\label{eq:membrane_A2}
u_m(t) = \sum_{i}^N w_{im} \tilde y_i(t) - \sum_{j\neq m} \beta \tilde z_j(t) +  \alpha - w_\text{norm}.
\end{equation}
Note that $w_\text{norm}$ can be incorporated into $\alpha$.
 Such constant weight sum could be enforced in a biological network by a synaptic scaling mechanism \cite{turrigiano2004homeostatic,savin2010independent} that normalizes the sum of incoming weights to a neuron. For the data-based model $\MD$, \cite{JonkeETAL:17a} used uniform excitabilities $\alpha$ for all excitatory neurons and no homeostasis for simplicity, see \me{membrane}.
We will argue below under which conditions this approximation is justified.
We consider the posterior distribution of such circuits as our second approximation of the exact posterior: 
\begin{align}\label{eq:post_A2}
p_\text{A2}(\v z| \v y, \v W) = \frac{1}{Z}\exp\left\{ \gamma \cdot \left( \sum_{i,m} w_{im} y_i z_m - \frac{1}{2} \sum_{m\neq l} \beta z_m z_l + \sum_m z_m (\alpha - w_\text{norm}) \right)    \right\}.
\end{align}
Note that in this case, $w_\text{norm}$ effectively leads to a smaller mean of the Gaussian prior. A detailed discussion about how parameters of the generative probabilistic model $\P$ can be mapped to parameters of the data-based microcircuit motif model $\MD$ is provided in {\em Network parameter interpretation} in {\em Methods}. 


\begin{figure}[tp]
\begin{center}
\includegraphics[width=0.6\textwidth]{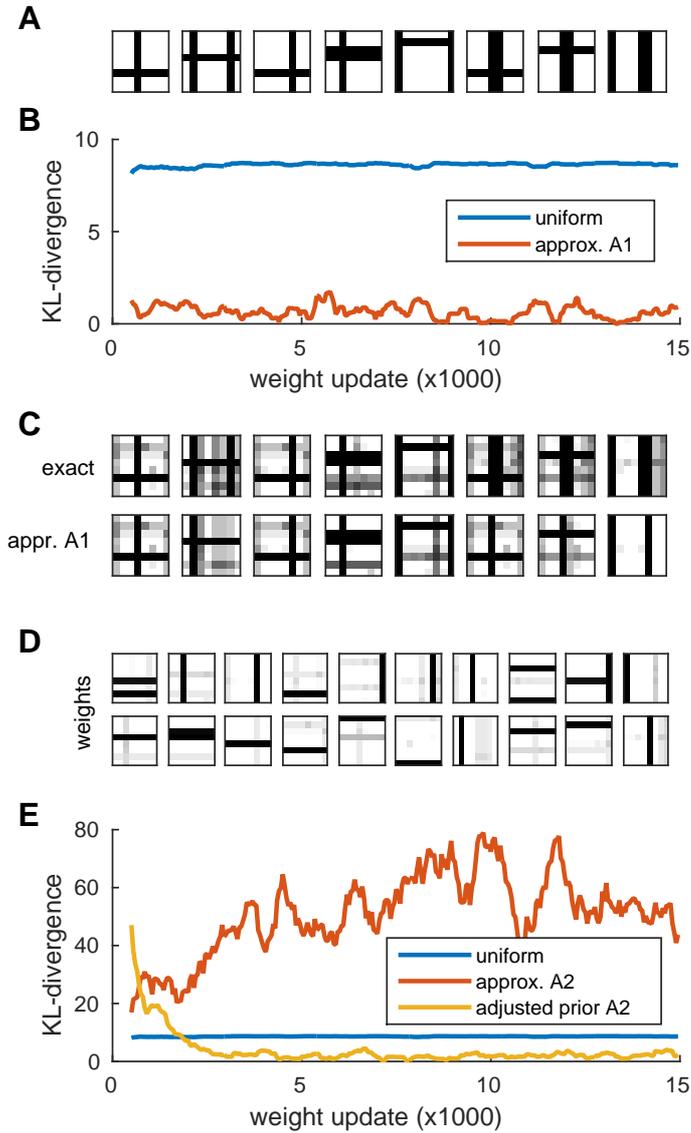}
\end{center}
\caption{{\bf Empirical evaluation of approximations in a superposition-of-bars task.}
\textbf{A}) Sample input patterns, depicted on the 8$\times$8 grid. Patterns consist of a varying number of superimposed horizontal or vertical bars. 
\textbf{B}) Evolution of the Kullback-Leibler divergence between the exact posterior and the posterior of approximation A1 during learning (red). 
As a comparison, the KL-divergence to a uniform distribution is indicated in blue. 
\textbf{C}) Example reconstruction of inputs from panel A from the posterior according to the hidden states with maximum probability in the exact posterior (top row) 
and the posterior of approximation A1 after learning (bottom). Scale between 0 (white) and 1 (black). 
\textbf{D}) Weights vectors of network neurons depicted on the 8$\times$8 grid as the input in panel A. Scale between 0 (white) and 6 (black). 
\textbf{E}) Evolution of the Kullback-Leibler divergence between the exact posterior, the posterior of approximation A2 during learning (red), 
and the posterior of approximation A2 with an adjusted sparsity prior d (yellow) during learning. As a comparison, the KL-divergence to a uniform distribution is indicated in blue.}
\label{fig:approx}
\end{figure}

We evaluated the impact of these two approximations in F\"oldiak's superposition-of-bars problem \cite{foldiak1990forming}. This is a standard blind-source separation problem that has also been used in \cite{JonkeETAL:17a} to evaluate the data-based network model $\MD$. In this problem, input patterns $\by$ are two-dimensional pixel arrays on which horizontal and vertical bars (lines) are superimposed, see Fig.~\ref{fig:approx}A. Input patterns were generated with a superposition of 1 to 3 bars from the distribution that was used in \cite{JonkeETAL:17a}. We performed inference of hidden causes by sampling from the approximate posterior distribution $p_\text{A1}(\v z| \v y, \v W)$ given by eq.~\eqref{eq:post_A1} for a network of 20 hidden-cause neurons. We performed approximate stochastic online EM in order to optimize the parameters of the model. The synaptic update rule \eqref{eq:update1} used for parameter updates is discussed in detail below. We compared the approximated posterior to the exact one \eqref{eq:post} by computing the Kullback-Leibler (KL) divergence $D_\text{KL}(p(\v z| \v y, \v W) || p_\text{A1}(\v z| \v y, \v W))$. The KL divergence was small throughout learning, with a slight decrease during the process, see Fig.~\ref{fig:approx}B (mean KL divergence during the second half of training was $0.55$). To evaluate what that KL-divergence means for the inference, we considered the hidden state vector $\v z_\text{max}$ with the maximum posterior probability after learning in the exact and approximate posterior and used this to reconstruct the input pattern by computing $\hat y = \sigma_\text{LS}(W^T \v z_\text{max})$. The reconstructed inputs for the 8 example inputs of Fig.~\ref{fig:approx}A are shown in Fig.~\ref{fig:approx}C for the exact posterior (top) and the approximate posterior (bottom). The approximate reconstructions resemble the exact ones in many cases, with occasional misses of a basic pattern. The final weights of the 20 neurons are shown in Fig.~\ref{fig:approx}D in the two-dimensional layout of the input to facilitate interpretability. Note that all basic patterns were represented by individual neurons with additional neurons that specialized on combined patterns.

The approximate posterior $p_\text{A2}$ is equivalent to the approximate posterior $p_\text{A1}$ if the synaptic weights of each neuron are normalized to a common $L1$ norm. It turned out in \cite{JonkeETAL:17a} that such a normalization is not strictly necessary. In this work, the data-based model $\MD$ managed to perform blind source separation with a posterior that can be best described by $p_\text{A2}$ without normalization of synaptic efficacies. We found that for the superposition-of-bars problem, the posterior distribution $p_\text{A2}$ differs significantly from the exact posterior if we set $w_\text{norm}=0$ in eq.~\eqref{eq:post_A2}, see red line in Fig.~\ref{fig:approx}E. This difference is mostly induced by the tendency of $p_\text{A2}$ to prefer many hidden causes due to the missing last term in eq.~\eqref{eq:post_A2}. If the prior was corrected to reduce the number of hidden causes, we found that the approximation was significantly improved, in particular as the network weight vectors approached their final norm values, see yellow line in Fig.~\ref{fig:approx}E. A closer inspection of eq.~\eqref{eq:post_A2} in comparison with eq.~\eqref{eq:post_A1} shows that this approximation is effective if basic patterns consist of a similar number of active units, because otherwise patterns with strong activity are preferred over weakly active ones (this is exactly what the last term in eq.~\eqref{eq:post_A1} compensates for). 

We conclude from this analysis that the microcircuit motif model $\MD$ approximates probabilistic inference in a noisy-OR-like probabilistic model of its inputs. The prior on network activity favors sparse network activity, but does not strictly enforce a predefined activity level. Such a more flexible regulation of network activity is obviously important when the network input is composed of a varying number of basic component patterns. 
We show below that this network behavior in combination with STDP allows the microcircuit motif model $\MD$ to perform blind source separation of mixed input sources.
Our analysis above has shown that the computation of blind source separation in $\MD$ can be facilitated either by the normalization of activity in input populations, or by homeostatic mechanisms that normalize excitatory synaptic efficacies within each neuron.

\subsubsection{STDP in the microcircuit motif model $\MD$ creates an internal model of network inputs:}

After we have established a link between a well-defined probabilistic model $\P$ and the spiking dynamics of microcircuit motif model $\MD$, we can now analyze plasticity in the network. 
The probabilistic model $\P$ defines a likelihood distribution over inputs $\by$ that depends on the parameters $\bW$ through
\begin{align}\label{eq:genmod_like}
p(\by|\bW) &= \sum_{\bz} p(\v z) p(\v y| \v z, \v W), 
\end{align}
where the sum runs over all possible hidden states $\bz$.

We propose that STDP in $\MD$ can be viewed as an adaptation of the parameters $\bW$ so that $p(\by|\bW)$ as defined by the probabilistic model $\P$ with parameters $\bW$ approximates the actually encountered distribution of spike inputs $p^*(\by)$ 
within the constraints of the prior $p(\bz)$. Since the prior typically is defined to favor sparse representations, this tends to extract the hidden sources of these patterns, an operation called blind source separation \cite{foldiak1990forming}.

More precisely, we show that STDP in $\MD$ approximates stochastic online EM \cite{Sato:99, Bishop:06} in $\P$. Given some external distribution $p^*(\v y)$ of synaptic inputs, EM adapts the model parameters $\v W$  such that the model likelihood distribution $p(\v y|\v W)$ approximates the given distribution $p^*(\v y)$.
More formally, the Kullback-Leibler divergence between the likelihood of inputs in the internal model $p(\by|\bW)$ and the empirical data distribution $p^*(\by)$ is brought to a local minimum.
The theoretically optimal learning rule for EM contains non-local terms which are hard to interpret from a biological point of view. 
In the following, we derive a local approximation to yield a simple STDP-like learning rule.

The goal of the EM algorithm is to find parameters that (locally) minimize the Kullback-Leibler divergence between the likelihood $p(\by|\bW)$ of inputs in the probabilistic model $\P$ and the empirical data distribution $p^*(\by)$, that is, the distribution of inputs experienced by the network $\MD$. This is equivalent to the maximization of the average data log-likelihood $E_{p^*}[\log p(\by|\bW)]$.
For a given set of training data $\bY$ and corresponding unobserved hidden variables $\bZ$, this corresponds to maximizing $\log p(\bY|\bW)$.
The optimization is done by iteratively performing two steps. For given parameters $\bWo$, the posterior distribution over hidden variables $p(\bZ|\bY, \bWo)$ is determined (the E-step). Using this distribution, one then performs the M-step where $E_{p(\bZ|\bY,\bWo)}[\log p(\bY,\bZ|\bW)]$ is maximized with respect to $\bW$ to obtain better parameters for the model. These steps are guaranteed to increase (if not already at a local optimum) a lower bound $\mathcal{L} = E_{p(\bZ|\bY,\bWo)} \left[ \log \frac{p(\bY, \bZ |\bW)}{p(\bY, \bZ |\bWo)}\right]$ on the data log likelihood, that is, $\mathcal{L}\le \log p(\bY|\bW)$.
These steps are iterated until convergence of parameters to a local optimum \cite{Bishop:06}. Computation of the M-step in the probabilistic model $\P$ is hard. In the generalized EM algorithm, the M-step is replaced by a procedure that just improves the parameters, without necessarily obtaining the optimal ones for a single M-step. This can be done for example by changing the parameters in the direction of the gradient 
\begin{equation}
  \Delta w_{im} \propto \frac{\partial}{\partial w_{im}} E_{p(\bZ|\bY,\bWo)}[\log p(\bY,\bZ|\bW)].
\end{equation}	
Since in our model, we assume that synaptic efficacy changes are instantaneous for each pre-post spike pair, we need to consider an online-version of the generalized EM algorithm. In stochastic online EM, for each data example $\by^{(k)}$, a sample $\bz^{(k)}$ from the posterior is drawn (the stochastic E-step) and parameters are changed according to this sample-pair.
As shown above, the $\MD$ network implements an approximation of the stochastic E-step.
In the M-step, each parameter $w_{im}$ is then updated in the direction of the gradient $\Delta w_{im} \propto \frac{\partial}{\partial w_{im}} \log p(\v y^{(k)}, \v z^{(k)}| \v W)$. As the prior $p(\v z)$ in $\P$ does not depend on $\v W$, this is equivalent to $\Delta w_{im} \propto \frac{\partial}{\partial w_{im}} \log p(\v y^{(k)} | \v z^{(k)}, \v W)$. 
For the likelihood given by eq.~\eqref{eq:genmod_likefact}, this derivative is given by 
\begin{equation}\label{eq:dw}
 \frac{\partial}{\partial w_{im}} \log p (\by^{(k)}|\bz^{(k)}, \v W) = 
 \gamma z_m \left( y_i - \frac{\exp(a_i)}{1+\exp(a_i)} \right) = \gamma z_m \left( y_i - \sigma_{LS}(a_i) \right), 
\end{equation}
where $\sigma_{LS}$ denotes the logistic sigmoid function. 
Hence, the synaptic update rule for weight $w_{im}$ is given by
\begin{equation} \label{eq:lr_nonlocal}
  \Delta w_{im} = \eta z_m \left( y_i - \sigma_{LS}(a_i)   \right),
\end{equation}
where $\eta>0$ is a learning rate.
This learning rule is not local as it requires information about the activation of all output neurons as well as values of all synaptic weights originating from input neuron $i$. In order to make this biologically plausible we approximate rule \eqref{eq:lr_nonlocal} by 
\begin{equation}\label{eq:update1}
  \Delta w_{im} =  \eta z_m \left( y_i - \sigma_{LS}(\gamma w_{im}) \right),
\end{equation}
This rule uses only locally available information at the synapse. What are the consequences of this approximation during learning? If only a single neuron in the network is active, then the approximation is exact. Otherwise, the approximation ignores what other neurons contribute to the explanation of input component $y_i$. This means that for $y_i=1$, the weight will further be increased even if $y_i=1$ is already fully explained by the network activity. For $y_i=0$, the decrease will in general be smaller than in the exact rule (since weights are non-negative). Note however that only the magnitudes of weigh changes are affected, but not which weights change and the sign of the change. Hence, we can conclude that the angle between the approximate parameter change vector and the exact parameter change vector is between $0$ and $90$ degrees. In other words, the inner product of these two vectors is always non-negative and the updates are performed in the correct direction. This was confirmed in simulations. In the learning experiment described in Fig.~\ref{fig:approx}, we compared the approximate update (that was used to optimize the model) with the update that was proposed by the exact rule \eqref{eq:dw} at every $50^\text{th}$ update step. The angle between the exact and approximate update vector was between $0^\circ$ and $84^\circ$ with a mean of $57^\circ$.

 In the simplified dynamics, a value of $z_m^{(k)}=1$ is indicated by a spike in network neuron $m$ when pattern $\v y^{(k)}$ is presented as input. We therefore map this update to the following synaptic plasticity rule: For each postsynaptic spike, update weight $w_{im}$ according to
\begin{align}\label{eq:STDP_meth}
\Delta w_{im} = \eta \left(y_i(t) - \sigma_{LS}(\gamma w_{im}) \right)\;.
\end{align}
According to this learning rule, when the presynaptic neuron $i$ spikes shortly before the postsynaptic neuron this results in long-term potentiation (LTP) which is weight dependent according to the term $\sigma_{LS}(w_{im})$. Due to the weight dependence, large weights lead to small weight changes, with vanishing changes for very large weights. When a post-synaptic spike by neuron $m$ is not preceded by a presynaptic spike by neuron $i$ (e.g.~when the pre-synaptic spike comes after the post-synaptic spike), this results in long term depression (LTD). LTD is also weight dependent, but to a much lesser extent as the weight-dependent factor varies only between $0.5$ and $1$. This behavior is mimicked by the standard STDP rule implemented in the data-based model $\MD$ that a standard weight dependence where updates exponentially decreased with $w_{im}$ for LTP and did not depend on $w_{im}$ for LTD. 

Hence, the dynamics and synaptic plasticity of the data-based model $\MD$ can be understood as an approximation of EM in the probabilistic model $\P$, that creates an internal model for the distribution $p^*(\by)$ of network inputs. This internal probabilistic model is defined by a noisy-OR-like likelihood term and a sparse prior on the hidden causes of the current input pattern. Hence, STDP can be understood as optimizing model parameters such that the observed distribution of input patterns can be explained through a set of basic patterns (hidden causes). It is assumed that the input at each time point can be described by a combination of a sparse subset of these patterns. In other words, STDP in the microcircuit motif model $\MD$ performs blind source separation of input patterns.

\section{Discussion}

We have provided a novel theoretical framework for analyzing and understanding computational properties that emerge from STDP in a prominent cortical microcircuit motif: interconnected populations of pyramidal cells and $\text{PV}^+$ interneurons in layer 2/3.
The computer simulations in \cite{JonkeETAL:17a}, that were based on the data from \cite{avermann2012microcircuits}, indicate that 
the computational operation of this network motif cannot be captured adequately by a WTA model.
Instead, this work suggests a k-WTA model, where a varying number of the most excited neurons become active. 
Since the WTA circuit model turns out to be inadequate for capturing the dynamics of interacting pyramidal cells and $\text{PV}^+$ interneurons, one needs to replace the probabilistic model that one had previously used to analyze the impact of STDP on the computational function of the network motif. 
Mixture models such as those proposed by \cite{nessler2013bayesian} and \cite{habenschuss2013emergence} are inseparably tied to WTA dynamics: For drawing a sample from a mixture model one first decides stochastically from which component of the mixture model this sample should be drawn (and only a single component can be selected for that). We have shown here that a quite different generative model, similar to the noisy-OR model, captures the impact of soft lateral inhibition on emergent network codes and computations much better (Fig.~\ref{fig:fig_generative_model}). The noisy-OR model is well-known in machine learning \cite{Neal:93,saund1995multiple}, but has apparently not previously been considered in computational neuroscience.
Our probabilistic model $\P$ further suggests that the varying number of active neurons in the circuit may depend both on a prior that is encoded by the network parameters and the familiarity of the network input.

We have shown that the evolution of the dynamics and computational function of the network motif under STDP can be understood from the theoretical perspective as an approximation of expectation maximization (EM) for fitting a noisy-OR based generative model to the statistics of the high dimensional spike input stream. This link to EM is very helpful from a theoretical perspective, since EM is one of the most useful theoretical principles that are known for understanding self-organization processes.
In particular, this theoretical framework allows us to elucidate emergent computational properties of the network motif for spike input streams that contain superimposed firing patterns from upstream networks. It disentangles these patterns and represents the occurrence of each pattern component by a separate sparse assembly of neurons, as already postulated in \cite{foldiak1990forming}. 

The established relationship between the network $\MD$ and the probabilistic model $\P$ allows us to relate the network parameters $\alpha$ and $w^{\text{IE}}$ (\me{membrane}) of $\MD$ to the parameters of the generative model $\P$. Briefly (for a detailed discussion, see {\em Network parameter interpretation} in {\em Methods}), the excitability $\alpha$ of pyramidal cells is proportional to $\frac{2 \mu -1}{2\sigma^2}$, see eq.~\eqref{eq:alpha}. The strength of inhibitory connections $w^{\text{IE}}$ to the pool of pyramidal cells is proportional to $\frac{1}{\sigma^2}$, see eq.~\eqref{eq:beta}. Hence, a large $\mu$ in combination with a small $\sigma^2$ (i.e., a sharp activity prior), leads to a large spontaneous activity that is tightly regulated by strong inhibitory feedback. On the other hand, a broad prior (larger $\sigma^2$) leads to weaker inhibitory feedback, thus allowing the network to attain a broader range of activities. 

\subsubsection*{Related work}

A related theoretical study for WTA circuits was performed in \cite{nessler2013bayesian, HabenschussETAL:13, kappel2014stdp} and extended to sheets of WTA circuits in \cite{bill2015distributed}. It was assumed in these models that inhibition normalizes network activity exactly, leading to a strict WTA behavior. The analysis in the present work is much more complex and necessarily has to include a number of approximations. Out analysis reveals that the softer type of inhibition that we studied provides the network with additional computational functionality.
There exists also a structural similarity of the proposed learning rule \eqref{eq:STDP_meth} to those reported in \cite{nessler2013bayesian, HabenschussETAL:13, kappel2014stdp}. This is insofar significant as it raises the question why the application of almost the same learning rule in one motif leads to learning and extraction of a single hidden cause and in another to the extraction of multiple causes. The answer most likely lies in the interplay between ``prior knowledge'' in the model (e.g.~in the form of the intrinsic excitability of neurons), the learning rule and inhibition strength: As there are multiple neurons in the proposed microcircuit motif model which can spike in response to the same input, each one of them can adapt its synaptic weights to increase the likelihood of spiking again whenever the same or a similar input pattern is presented in the future, possibly in conjunction with other different input components. This is manifested through increased total input strength to those neurons when the pattern is seen again. But this results also in increased total inhibition to all other neurons, thereby effectively limiting the number of winners. As there is no fixed normalization of firing rates (probabilities), as soon as the input strength caused by a single feature component is strong enough to trigger the spike in some neuron, the neuron will respond to each pattern which consists of that particular feature. On average this will force neurons to specialize on a single feature component. Therefore, after learning, each spike can be interpreted as indication of a particular feature component.

The noisy-OR model \me{nOR} is tightly related to the likelihood model used in this article. It is one of the most basic likelihood models that allows to combine basic patterns. Noisy-OR and related models have previously been used in the machine learning literature as models for nonlinear component extraction \cite{saund1995multiple,lucke2008maximal}, or as basic elements in belief networks \cite{neal1992connectionist}, but they have so far not been linked to cortical processing.

The extraction of reoccurring components of input patterns is closely related to blind source separation and independent component analysis (ICA) \cite{HyvaerinenETAL:04}. Previous work in this direction includes implementations of ICA in artificial neural networks \cite{hyvarinen1999fast}, see also \cite{lucke2010expectation}. These abstract models are only loosely connected to computation in cortical network motifs.
\cite{savin2010independent} investigated ICA in the context of spiking neurons. Theoretical rules for intrinsic plasticity were derived which enable neurons in combination with input normalization, weight scaling, and STDP, to extract independent components of inputs. An interesting difference is that the inhibition in \cite{savin2010independent} acts to decorrelate neuronal activity. Intrinsic plasticity on the other hand enforces sparse activity (this sparsening has to happen on the time scale of input presentations). In our probabilistic model $\P$, sparse network activity is enforced by a prior over network activities, implemented in $\MD$  through the inhibitory feedback that models experimentally found network connectivity \cite{avermann2012microcircuits}. This inhibition naturally acts on a fast time scale \cite{OkunLampl:08}, while the time scale for intrinsic plasticity is unclear \cite{turrigiano2004homeostatic}. 

\section{Methods}\label{sec:methods}

\subsection{Definition of $\boldsymbol{\MD}$: Data-based network model for a layer 2/3 microcircuit motif}
The layer 2/3 microcircuit motif was modeled in \cite{JonkeETAL:17a} by the data-based model $\MD$. The model is described here briefly for completeness. See  \cite{JonkeETAL:17a} for a thorough definition. The model $\MD$ consists of 
two reciprocally connected pools of neurons, an excitatory pool and an inhibitory pool.  Inhibitory network neurons are recurrently connected. Excitatory network neurons receive additional excitatory synaptic input from a pool of $N$ input neurons.
Fig.~\ref{fig:fig_model}A summarizes the connectivity structure of the data-based model $\MD$ together with connection probabilities.
Connection probabilities have been chosen according to the experimental data described in \cite{avermann2012microcircuits}.


Let $t_i^{(1)}, t_i^{(2)}, \dots$ denote the spike times of input neuron $i$.
The {\em output trace} $\tilde y_i(t)$ of input neuron $i$ is given by the temporal sum of unweighted postsynaptic potentials (PSPs) arising from input neuron $i$:   
\begin{align} \label{eq:output_trace}
	\tilde y_i(t) = \sum_{f} \epsilon(t-t_i^{(f)}),
\end{align}
where $\epsilon$ is the synaptic response kernel, i.e., the shape of the PSP. 
It is given by a double-exponential function with a rise time constant $\tau_r=1$ ms and a fall time constant $\tau_f=10$ ms.
For given spike times, output traces of excitatory network neurons and inhibitory network neurons are defined analogously and denoted by $\tilde z_m(t)$ and $I_j(t)$ respectively.

The network consists of $M=400$ excitatory neurons, modeled as stochastic spike response model neurons \cite{JolivetETAL:06}, see eqs.~\eqref{eq:neuron_firing_prob} and \eqref{eq:membrane}. 
See Sec.~\ref{sec:param_mot} for a motivation of network parameter values from a theoretical perspective.

The instantaneous firing rate $\rho_m$ of neuron $m$ can be re-written (by substituting \me{membrane} in \me{neuron_firing_prob}) as: 
\begin{align}\label{eq:neuron_firing_prob_divisive}
  \rho_m(t) &= 
  \frac{1}{\tau} \frac{\exp\left(\gamma \sum_i w_{im} \tilde y_i(t) + \gamma \alpha \right)}{\exp\left(\gamma \sum_{j \in \mathcal{I}_m} w^{\text{IE}} I_j(t)\right)}\;\;.
\end{align}
Here, the numerator includes all excitatory contributions to the firing rate $\rho_m(t)$. The denominator in this term for the firing rate describes inhibitory contributions, thereby reflecting divisive inhibition \cite{CarandiniHeeger:12}.

Apart from excitatory neurons there are $M_{\text{inh}}=100$ inhibitory neurons in the network. 
Inhibitory neurons are also modeled as stochastic spike response neurons with an instantaneous firing rate given by 
\begin{align}\label{eq:inh_firing_prob}
  \rho_m^{\text{inh}}(t)\ &= \sigma_\text{rect}(u_m^{\text{inh}}(t)),
\end{align}
where $\sigma_\text{rect}$ denotes the linear rectifying function $\sigma_\text{rect}(u)=u$ for $u\ge 0$ and $0$ otherwise. The membrane potentials of inhibitory neurons are given by
\begin{align}
u_m^{\text{inh}} (t) = \sum_{i \in \mathcal{E}_m} w^{\text{EI}} \tilde z_i(t) -\sum_{j \in \mathcal{II}_m} w^{\text{II}} I_j(t),\label{eq:inh_membrane} 
\end{align}
where $\tilde z_i(t)$ denotes synaptic input (output trace) from excitatory network neuron $i$,
$\mathcal{E}_m$ ($\mathcal{II}_m$) denotes the set of indices of excitatory (inhibitory) neurons that project to inhibitory neuron $m$, and
$w^{\text{EI}}$ ($w^{\text{II}}$) denotes the excitatory (inhibitory) weight to inhibitory neurons.

Synaptic connections from input neurons to excitatory network neurons are subject to STDP. A standard version of STDP is employed with an exponential weight dependency for potentiation \cite{habenschuss2013emergence}, see \cite{JonkeETAL:17a}. 

The simulations for Fig.~\ref{fig:fig_tilted} are described in detail in \cite{JonkeETAL:17a}.

\subsection{Network parameter interpretation: }\label{sec:param_mot}

In the section {\em Interpretation of the dynamics of $\boldsymbol{\MD}$ in the light of $\boldsymbol{\P}$}, we have established a relationship between the parameters of the probabilistic model $\P$ and network parameters. This relationship was however derived based on a simplified network model that included for example rectangular EPSPs and direct inhibitory connections without explicit inhibitory neurons. Nevertheless, one can also determine reasonable parameter settings for the data-based model $\MD$ based on a prior on network activity that is defined in the probabilistic model $\P$. These parameters are the excitability $\alpha$ and the synaptic weights between and within excitatory and inhibitory network neurons.

In this section we start by assuming such a prior \me{genmod_prior} with parameters $\mu=-3.4$ 
(this includes already a correction of the prior for the missing $w_\text{norm}$) and $\sigma^2=0.35$ 
as well as a fitting parameter $\gamma=2$ (eq.~\ref{eq:neuron_firing_prob}) and deduce the parameters used in the simulations. As shown above, the neural excitability $\alpha$ is then given by 
$\alpha=\frac{1}{\gamma} \frac{2\mu-1}{2\sigma^2}$. 
We obtain for the chosen $\gamma=2$:
\begin{equation}\label{eq:alpha}
  \alpha = \frac{1}{\gamma} \left(\frac{2\mu-1}{2\sigma^2}\right) = -5.57.
\end{equation}
The inhibition strength $\beta$ of the approximate dynamics \me{membrane_NS} is replaced by the weight $w^\text{IE}$ from inhibitory neurons to excitatory neurons in $\MD$. 
From the probabilistic model, we determined $\beta$ as $\beta=\frac{1}{\gamma \sigma^2}$ under the assumption of rectangular inhibitory PSPs. In $\MD$ we use double-exponential PSPs instead of rectangular ones. One therefore has to correct for differences in the PSP integrals. Using this correction, one obtains   
\begin{equation}
  \beta'= c_{\text{PSP}} \beta = \frac{c_{\text{PSP}}}{\gamma \sigma^2} = 1.114,
\end{equation}
where $c_{\text{PSP}}$ is the ratio between the integrals over the rectangular PSPs used in the approximate dynamics and the double-exponential PSPs used in $\MD$ ($c_{\text{PSP}}=0.78$ for the shapes used for $\MD$).

The weights $w^\text{IE}$ can be determined by comparing \me{membrane} with \me{membrane_NS} with corrected inhibition strength 
\begin{equation}
  \sum_{j \in \mathcal{I}_m} w^{\text{IE}} I_j(t) = \sum_{j=1}^M \beta' z_j(t).
\end{equation}  
Under the assumption that the number of spikes in the pool of inhibitory neurons is at any time (with a slight delay) approximately equal to the number of spikes in the pool of excitatory neurons, we obtain
\begin{equation}
  p^{\text{IE}} w^{\text{IE}} = \beta',
\end{equation}  
where $p^{\text{IE}}$ denotes the connection probability from inhibitory to excitatory network neurons. This yields
\begin{equation}\label{eq:beta}
  w^{\text{IE}} = \beta' \frac{1}{p^{\text{IE}} } = \frac{c_{\text{PSP}}}{\gamma \sigma^2 p^{\text{IE}} } = 1.86.
\end{equation}  
We now first consider the weights $w^{\text{EI}}$ from excitatory to inhibitory neurons under the assumption of no I-to-I connections. In this case, in order to obtain the same number of spikes in the inhibitory neurons as in the excitatory neurons, each spike from an excitatory neuron should induce on average one spike within all inhibitory neurons, that is, 
\begin{equation}
  w^{\text{EI,no II}} \bar \epsilon M_\text{inh} p^{\text{EI}} = 1,
\end{equation}  
where $\bar \epsilon=0.01/c_\text{PSP}$ is the integral over the alpha-PSPs, $p^{\text{EI}}$ is the connection probability from excitatory to inhibitory neurons, and $M_\text{inh}=100$ is the number of inhibitory neurons. We obtain
\begin{equation}
  w^{\text{EI,no II}} = c_\text{PSP}/p^{\text{EI}} = 1.357.
\end{equation}  
Without I-to-I connections, this guarantees that excitation is balanced by inhibition. However, the single spike (on average) will occur on average with a delay of $5$ ms. Interestingly, the I-to-I connections can help to decrease this delay. In particular, if one demands that the inhibitory spike is elicited with a delay of less than one ms on average, then one can simply increase the weights $w^{\text{EI}}$ by some factor $c^\text{EI}=10$, leading to  
\begin{equation}
  w^\text{EI} = w^{\text{EI,no II}} c^\text{EI} = c_\text{PSP} c^\text{EI} / p^{\text{EI}} = 13.57.
\end{equation}  
Now, each spike in the excitatory population induces in the inhibitory population for approximately $10$ ms a total rate of $1000$ Hz, leading to an average delay of $1$ ms. Without I-to-I connections this would however lead to too many successive spikes within these $10$ ms. The I-to-I connections can compensate this too large excitation. For an approximately correct compensation, the first inhibitory spike has to balance out this excitation, which is approximately achieved by providing exactly the same amount of inhibition to inhibitory neurons, leading to 
\begin{equation}
  w^\text{II} = c_\text{PSP} c^\text{EI} / p^{\text{II}} .
\end{equation}  
Since $p^{\text{II}} \approx p^{\text{EI}}$, we used $w^\text{II}=w^\text{EI}$ for simplicity.
These are the parameter values used for the $\MD$ model in \cite{JonkeETAL:17a}.

\subsection{Details to simulations for Figure \ref{fig:approx} }\label{sec:details_approx}

{\bf Creation of basic patterns:}
Basic patterns were $64$-dimensional vectors, each representing a horizontal or vertical bar on an $8 \times 8$ two-dimensional pixel array. We defined $16$ basic patterns $\v x^{(1)}, \dots, \v x^{(16)}$ in total, corresponding to all possible horizontal and vertical bars of width $1$ in this pixel array. For a horizontal (vertical) bar, all pixels of a row (column) in the array attained the value $1$ while all other pixels were set to $0$. The entries of the basic pattern vectors were then defined by the values of the corresponding pixels in the array.

\vspace{0.2cm}
\noindent
{\bf Superposition of basic patterns:}
To generate an input pattern, basic rate patterns were superimposed as follows. The number of superimposed basic patters $n_{sup}$ was chosen between $1$ and $3$ drawn from the distribution $p(n_{sup}=k) = \frac{0.9^k 0.1^{3-k}}{\sum_{l=1}^{3}0.9^l 0.1^{3-l}}$.  Then each basic pattern to be superimposed was drawn uniformly from the set of basic patterns without replacement. This corresponds to the distribution used in \cite{JonkeETAL:17a}. The input vector $\v y$ was then given by $\v y = \max\{1, \sum_{i=1}^{n_{sup}} \v x^{(bp(i))}\}$, where $bp(i)$ denotes the $i^\text{th}$ basic pattern to be superimposed and the $\max$ operation is performed element-wise.

\vspace{0.2cm}
\noindent
{\bf Optimization of the generative model:}
The generative model \eqref{eq:genmod_prior}--\eqref{eq:genmod_likefact} with $20$ hidden causes $\v z$ was fitted to this data in an iterative manner. One iteration of the fitting algorithm was performed as follows:
\begin{enumerate}
  \item draw an input vector $\v y$ as described above;
  \item draw a sample from the approximate posterior $p_\text{A1}(\v z| \v y, \v W)$, eq.~\eqref{eq:post_A1};
  \item update the parameters $\v W$ of the model according to eq.~\eqref{eq:update1}.
\end{enumerate}
Since the posterior in step (2) is intractable, it was approximated by assuming that a maximum of $4$ hidden causes are active in the posterior distribution (state vectors with more active hidden causes usually had negligible probabilities). This allowed us to compute the partition function and therefore to sample hidden state vectors in a straight-forward manner. Further, we did not consider hidden state vectors with no active hidden state since those would not lead to any parameter changes.

\vspace{0.2cm}
\noindent
{\bf Parameters of the model and learning rule:} 
A prior distribution $p(\v z)$ with parameters $\mu=6$ and $\sigma^2=0.35$ was used. 
The scaling parameter $\gamma$ was set to $1$. 
Weights $w_{ij}$ were initialized with values drawn from a uniform distribution in $[0,0.1]$. Weights were clipped between a minimal value of $0$ and a maximal value of $6$. A constant learning rate of $\eta=0.1$ was used. Training was performed for $15000$ updates. Network characteristics (such as KL divergences) were computed every $50^\text{th}$ update.

\vspace{0.2cm}
\noindent
{\bf Figure \ref{fig:approx}B:} 
Every $50^\text{th}$ update, we computed the KL-divergence between the true posterior and the approximate posterior $D_\text{KL}(p(\v z| \v y, \v W) || p_\text{A1}(\v z| \v y, \v W))$. In addition we also computed the KL-divergence between $p(\v z| \v y, \v W)$ and a uniform distribution over state vectors. In all these divergences, we only considered the distribution over vectors with at most $4$ active hidden causes for tractability (see above). For Fig.~\ref{fig:approx}B, the data was smoothed using a box-car filter of size $10$. 

\vspace{0.2cm}
\noindent
{\bf Figure \ref{fig:approx}C:} 
We considered the input patters given in panel A and computed the hidden state $\v z_\text{max}$ with the maximum posterior probability (computed as described above) after learning in the exact and approximate posterior. This hidden state vector was then used to reconstruct the input pattern by computing $\hat y = \sigma_\text{LS}(W^T \v z_\text{max})$. Note that this is not a sample of $\v y$ but it defines the probability of each individual pixel to be $1$. 

\vspace{0.2cm}
\noindent
{\bf Figure \ref{fig:approx}E:} 
Every $50^\text{th}$ update of the simulation described above, we computed the KL-divergence $D_\text{KL}(p(\v z| \v y, \v W) || p_\text{A2}(\v z| \v y, \v W))$ between the true posterior and the posterior according to approximation $A2$. For the red curve we used $p_\text{A2}$ with $w_\text{norm}=0$ and the same $\mu=6$ as given for the original model. For the yellow curve, we corrected the prior of the model to have $\mu=-12$. The KL-divergence to the uniform distribution was computed as described above.

\vspace{2cm}
\noindent
{\em Acknowledgements:} Written under partial support by the Human Brain Project of the European Union \#604102 and \#720270, and the Austrian Science Fund (FWF): I 3251-N33.

\bibliographystyle{apalike}

\end{document}